\PassOptionsToPackage{unicode}{hyperref}
\PassOptionsToPackage{hyphens}{url}
\PassOptionsToPackage{dvipsnames,svgnames,x11names}{xcolor}
\documentclass[
  12pt]{article}

\usepackage{amsmath,amssymb}
\usepackage{iftex}
\ifPDFTeX
  \usepackage[T1]{fontenc}
  \usepackage[utf8]{inputenc}
  \usepackage{textcomp} 
\else 
  \usepackage{unicode-math}
  \defaultfontfeatures{Scale=MatchLowercase}
  \defaultfontfeatures[\rmfamily]{Ligatures=TeX,Scale=1}
\fi
\usepackage{lmodern}
\ifPDFTeX\else  
\fi
\IfFileExists{upquote.sty}{\usepackage{upquote}}{}
\IfFileExists{microtype.sty}{
  \usepackage[]{microtype}
  \UseMicrotypeSet[protrusion]{basicmath} 
}{}
\makeatletter
\@ifundefined{KOMAClassName}{
  \IfFileExists{parskip.sty}{%
    \usepackage{parskip}
  }{
    \setlength{\parindent}{0pt}
    \setlength{\parskip}{6pt plus 2pt minus 1pt}}
}{
  \KOMAoptions{parskip=half}}
\makeatother
\usepackage{xcolor}
\makeatletter
\ifx\paragraph\undefined\else
  \let\oldparagraph\paragraph
  \renewcommand{\paragraph}{
    \@ifstar
      \xxxParagraphStar
      \xxxParagraphNoStar
  }
  \newcommand{\xxxParagraphStar}[1]{\oldparagraph*{#1}\mbox{}}
  \newcommand{\xxxParagraphNoStar}[1]{\oldparagraph{#1}\mbox{}}
\fi
\ifx\subparagraph\undefined\else
  \let\oldsubparagraph\subparagraph
  \renewcommand{\subparagraph}{
    \@ifstar
      \xxxSubParagraphStar
      \xxxSubParagraphNoStar
  }
  \newcommand{\xxxSubParagraphStar}[1]{\oldsubparagraph*{#1}\mbox{}}
  \newcommand{\xxxSubParagraphNoStar}[1]{\oldsubparagraph{#1}\mbox{}}
\fi
\makeatother

\usepackage{longtable,booktabs,array}
\usepackage{calc} 
\usepackage{etoolbox}
\makeatletter
\patchcmd\longtable{\par}{\if@noskipsec\mbox{}\fi\par}{}{}
\makeatother
\IfFileExists{footnotehyper.sty}{\usepackage{footnotehyper}}{\usepackage{footnote}}
\makesavenoteenv{longtable}
\usepackage{graphicx}
\makeatletter
\newsavebox\pandoc@box
\newcommand*\pandocbounded[1]{
  \sbox\pandoc@box{#1}%
  \Gscale@div\@tempa{\textheight}{\dimexpr\ht\pandoc@box+\dp\pandoc@box\relax}%
  \Gscale@div\@tempb{\linewidth}{\wd\pandoc@box}%
  \ifdim\@tempb\p@<\@tempa\p@\let\@tempa\@tempb\fi
  \ifdim\@tempa\p@<\p@\scalebox{\@tempa}{\usebox\pandoc@box}%
  \else\usebox{\pandoc@box}%
  \fi%
}
\def\fps@figure{htbp}
\makeatother

\usepackage{booktabs}
\usepackage{longtable}
\usepackage{array}
\usepackage{multirow}
\usepackage{wrapfig}
\usepackage{float}
\usepackage{colortbl}
\usepackage{pdflscape}
\usepackage{tabu}
\usepackage{threeparttable}
\usepackage{threeparttablex}
\usepackage[normalem]{ulem}
\usepackage{makecell}
\usepackage{xcolor}
\makeatletter
\@ifpackageloaded{caption}{}{\usepackage{caption}}
\AtBeginDocument{%
\ifdefined\contentsname
  \renewcommand*\contentsname{Table of contents}
\else
  \newcommand\contentsname{Table of contents}
\fi
\ifdefined\listfigurename
  \renewcommand*\listfigurename{List of Figures}
\else
  \newcommand\listfigurename{List of Figures}
\fi
\ifdefined\listtablename
  \renewcommand*\listtablename{List of Tables}
\else
  \newcommand\listtablename{List of Tables}
\fi
\ifdefined\figurename
  \renewcommand*\figurename{Figure}
\else
  \newcommand\figurename{Figure}
\fi
\ifdefined\tablename
  \renewcommand*\tablename{Table}
\else
  \newcommand\tablename{Table}
\fi
}
\@ifpackageloaded{float}{}{\usepackage{float}}
\floatstyle{ruled}
\@ifundefined{c@chapter}{\newfloat{codelisting}{h}{lop}}{\newfloat{codelisting}{h}{lop}[chapter]}
\floatname{codelisting}{Listing}

\makeatother
\makeatletter
\@ifpackageloaded{caption}{}{\usepackage{caption}}
\@ifpackageloaded{subcaption}{}{\usepackage{subcaption}}
\makeatother

\usepackage[]{natbib}
\usepackage{bookmark}

\IfFileExists{xurl.sty}{\usepackage{xurl}}{} 
\hypersetup{
  pdftitle={Statisticians Training STEM Educators in Statistics Methods and Pedagogy: A Case Study of Instructor Training in Bayesian Methods},
  pdfauthor={Mine Dogucu; Jingchen Hu; Amy H Herring},
  pdfkeywords={faculty professional development, statistics
education, data science education, STEM education, train the trainer
(TTT), faculty pedagogical training},
  colorlinks=true,
  linkcolor={blue},
  filecolor={Maroon},
  citecolor={Blue},
  urlcolor={Blue},
  pdfcreator={LaTeX via pandoc}}

\begin{document}

\def\spacingset#1{\renewcommand{\baselinestretch}%
{#1}\small\normalsize} \spacingset{1}


\date{May 4, 2025}
\title{\bf Statisticians Training STEM Educators in Statistics Methods
and Pedagogy: A Case Study of Instructor Training in Bayesian Methods}
\author{
Mine Dogucu\thanks{All authors were supported by National Science
Foundation Improving Undergraduate STEM Education program with award
numbers 2215879, 2215920, and 2215709. Authors thank Postsecondary
Education Research \& Implementation Institute at University of
California Irvine for their support in evaluating the program.}\\
Department of Statistics, University of California, Irvine\\
and\\Jingchen Hu\\
Department of Mathematics and Statistics, Vassar College\\
and\\Amy H Herring\\
Department of Statistical Science, Duke University\\
}
\maketitle

\bigskip
\bigskip
\begin{abstract}
Educating the next generation of scientists in statistical methodology
is an important task. Educating their instructors in statistical content
knowledge and pedagogical knowledge is as important and provides an
indirect impact of students' learning. Statisticians are in a place to
lead train-the-trainer (TTT) programs in different methods. We present
our instructor training program in Bayesian methods as an effective case
study of a TTT model. In addition to describing the details of the
structure of our training program, we share our experience in designing
and implementing our program including the challenges we face, the
opportunities created, and our recommendations for TTT programs led by
statisticians.
\end{abstract}

\noindent%
{\it Keywords:} faculty professional development, statistics
education, data science education, STEM education, train the trainer
(TTT), faculty pedagogical training
\vfill

\newpage
\spacingset{1.9} 

\newcommand{\website}[1]{https://www.stat.uci.edu/bayes-bats/}
\newcommand{\programlong}[1]{BATS to enhance \textbf{Ba}yesian \textbf{T}hinking in \textbf{S}TEM}
\newcommand{\taspaper}[1]{\citep{DogucuHu2022TAS}}
\newcommand{\bats}[1]{BATS}
\newcommand{\UCI}[1]{University of California, Irvine}
\newcommand{\Vassar}[1]{Vassar College}

\section{Introduction}\label{sec-intro}

Educating the next generation of scientists in statistical methodology
is an important task. Scholars such as Gelman and Loken
(\citeyear{gelman2014statistical}) and Cox and Efron
(\citeyear{cox2017statistical}) have called for improved statistical
education across the sciences to enhance scientific findings. For some
scientists, training in statistical methods starts in an introductory
statistics course, which may or may not be taught by a statistician. In
other cases, exposure to statistics for STEM (Science, Technology,
Engineering, and Mathematics) students could begin and continue in their
disciplinary departments and programs, where courses are more likely be
taught by a non-statistician scientist. Therefore, it is important for
statisticians to train the educators of the next generation of
scientists in statistical methods. A Train-the-trainer (TTT) approach
provides an indirect impact on students' learning. This is especially
important for modern statistical methods, which non-statistician STEM
faculty may not have had exposure to yet during their own disciplinary
training and career.

TTT is a commonly used approach across multiple disciplines, including
public health \citep{orfaly2005train}, sustainability
\citep{cheung2018train}, and archaeology \citep{guinness2014they}, to
name a few. The TTT approach is also embraced in the broader statistics
community and the statistics education community. For instance, TTT is
used in interviewer training, which is essential for data collection in
large scale survey design \citep{alcser2005share}.

Many professional development programs of statistics and data science
education for teachers or college instructors can also be considered
TTT. For example, workshops at the United States Conference on Teaching
Statistics (USCOTS) span across statistical content including technology
as well pedagogy. Other examples include teacher training workshops such
as inSTEP, an online training platform in statistics and data science
aimed at teachers in grades 6-12 \citep{lee2023online}.

TTT programs can range from a few hours long workshops to multi-year
long programs. Regardless of the focus, discipline, and the length of a
TTT program, one common ground is that all TTT programs want to create
change, and they aim to do it through training the trainer. Similar to
many TTT programs, we want to create change - more precisely to create
more exposure to Bayesian methods for undergraduate STEM students. Over
the last three years we have run a faculty training program called
BATS to enhance \textbf{Ba}yesian \textbf{T}hinking in \textbf{S}TEM
funded by the National Science Foundation's (NSF's) Improving
Undergraduate STEM Education (IUSE) program. All authors of this
manuscript have served as Principal Investigators (PIs) on this
cross-institutional collaborative grant.

In writing this manuscript, we first aim at sharing an overview of our
program that we have run as a case study of the TTT approach. We believe
that our case study can potentially serve as an example for
statisticians who plan to design and deliver TTT programs in specific
statistical methods (e.g., time-series analysis, spatial analysis).
Second, we want to share our experience in implementing our TTT program
including the challenges we faced, the opportunities created, and our
recommendations for similar programs. Although our program focuses on
Bayesian methods, in this manuscript we do not intend to share content
on Bayesian methods or Bayesian pedagogy. Readers who are interested in
the Bayesian content and pedagogy can find our program materials online
at https://www.stat.uci.edu/bayes-bats/.

We describe the elements of our program in Section~\ref{sec-overview}.
Section~\ref{sec-outcomes} shares information about the demographics of
the participants and their institutions, as well as the impact created
as a result of our program. Lastly, we share recommendations and a few
other considerations for statisticians who might run their own TTT
programs in Section~\ref{sec-discussion}.

\section{Program Overview}\label{sec-overview}

\begin{figure}

\centering{

\pandocbounded{\includegraphics[keepaspectratio]{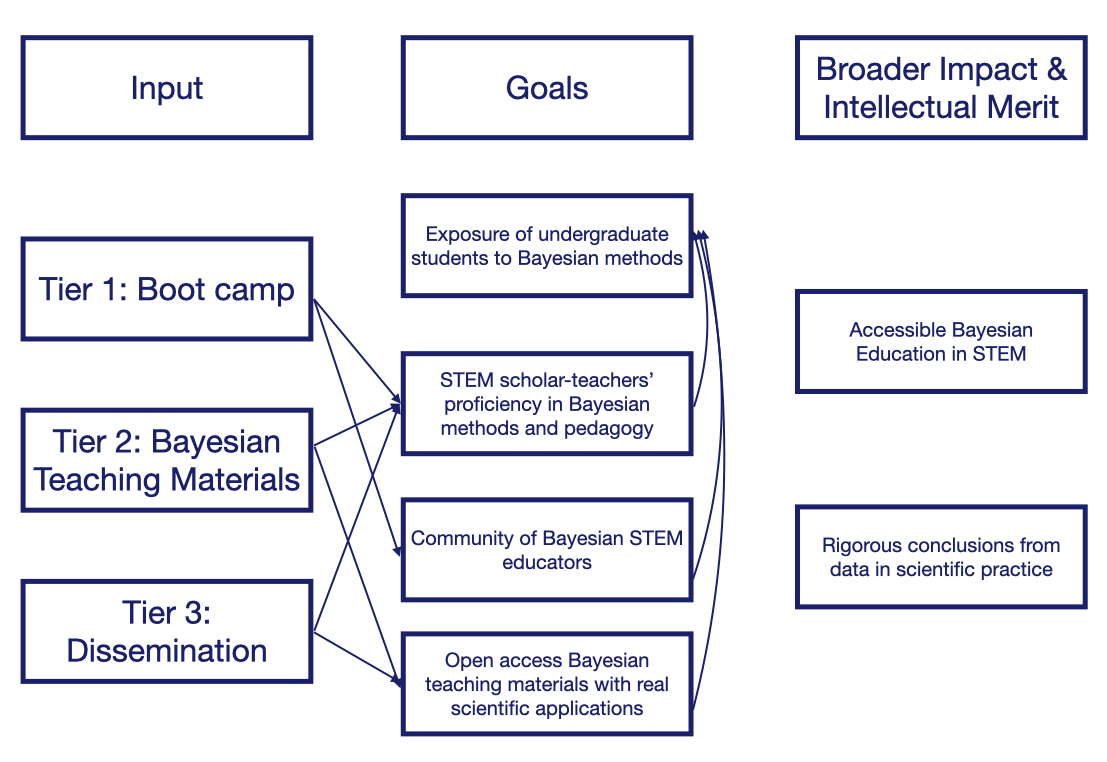}}

}

\caption{\label{fig-theoryofchange}Theory of Change Model of Our TTT
Program}

\end{figure}

As mentioned briefly in Section~\ref{sec-intro}, the ultimate goal of
our program is to increase exposure of undergraduate STEM students to
Bayesian methods. We further created three sub goals that we believe all
contribute to the ultimate goal. Based on the TTT approach, the first
sub goal is that we should enhance and improve STEM scholar-teachers'
proficiency in Bayesian methods in pedagogy. The second sub goal is to
create and foster a community of Bayesian STEM educators. The third and
final sub goal is to disseminate Bayesian teaching materials with real
scientific applications. Figure~\ref{fig-theoryofchange} illustrates the
ultimate goal and our three sub goals in the middle ``Goals'' column.
Some NSF IUSE proposals are required to utilize a theory of change model
\citep{connolly2015theories}, and this figure comes from our proposal,
which illustrates our theory of change model.

To achieve the ultimate goal and three sub goals, we designed a
three-tier structure for our program: Tier 1, a week-long summer
bootcamp to train a cohort of STEM educators in Bayesian methods; Tier
2, a semester-long mentoring experience for selected participants to
create their own course materials of Bayesian methods; and Tier 3,
support for selected participants to disseminate their course materials
through conference presentations and journal publications. As shown in
Figure~\ref{fig-theoryofchange} first ``Input'' column, each tier
contributes to two sub goals, and all tiers enhance and reinforce each
other. For example, Tier 3 dissemination directly facilitates the
creation of open access Bayesian teaching materials with real scientific
applications (sub goal 3), which in turn indirectly enhances and
improves STEM scholar-teachers' proficiency in Bayesian methods in
pedagogy (sub goal 1).

With all input and goals in place and properly executed, we hoped to
achieve accessible Bayesian education in STEM and draw rigorous
conclusions from data in scientific practice, shown in
Figure~\ref{fig-theoryofchange} third ``Broader Impact and Intellectual
Merit'' column.

In the past three years, we have trained two cohorts of program
participants. For Cohort 1, the Tier 1 bootcamp took place at
University of California, Irvine on the West Coast in July 2023; Tier 2
course materials development was from September through December 2023;
and Tier 3 dissemination started in January 2024 and is ongoing. For
Cohort 2, Tier 1 was at Vassar College on the East Coast in July 2024;
Tier 2 started in September and ended in December 2024; and Tier 3
started in January 2025 and is ongoing.

Next we dive deeper into each tier of our program in detail. In
describing each tier, we start with the content, followed by the
personnel and the participant support provided. We end with detailing
the recruitment process.

\subsection{Tier 1: Bootcamp}\label{tier-1-bootcamp}

Tier 1 is a week-long summer bootcamp to train STEM instructors in
Bayesian methods and pedagogy. Our bootcamp had lectures in the morning
and discussions and activities in the afternoon for five days.
Breakfast, lunch, and two coffee breaks were offered onsite as part of
our community building efforts. Before, during, and after the bootcamp,
a Slack group was used for communications and community building.

The bootcamp lecture topics were selected following recommendations from
\citep{DogucuHu2022TAS}: foundations of Bayesian inference, Bayesian
computing, and Bayesian modeling. Discussion topics were focused on
pedagogy and best practices; example topics include challenges of
teaching Bayes and effective assessments for Bayes learning. Activities
were hands-on experience for instructor participants to practice and
design Bayesian teaching materials, such as designing a Bayes theorem
activity and designing a posterior analysis activity. Some of these
practices naturally had grown into a Tier 2 project for an individual or
a team to spend a semester creating Bayesian teaching materials with
program organizers' mentoring. Prior to the bootcamp, we provided
virtual and asynchronous training in computing and probability and
statistics content.

All three authors served as instructors for the bootcamp, along with a
student teaching assistant. Every Tier 1 participant received a stipend
to help cover their travel and lodging for attending the bootcamp.

The call for Tier 1 applications was shared within academic and
professional networks of a number of STEM fields. In the application
form, each applicant was asked to share a list of courses taught and
enrollments over the past two years; their past exposure or skills in
statistics; statistical programming, and Bayesian inference, the way
they incorporated statistical and quantitative methods into their
teaching; what they hoped to gain from the bootcamp; and whether they
had any publications focused on pedagogical practice and their interest
in co-authoring such papers, among other things. The selection process
reviewed applicants' responses to these questions, while taking into
account of their institution types and position types to strike a
balance of recruiting tenure track and non-tenure track (including
lectures and teaching professors) participants from a variety of
institutions (universities, 4-year colleges, and community colleges). In
both program cohorts, we aimed at recruiting 20 participants each year,
and eventually 35 were recruited for Tier 1.

\subsection{Tier 2: Course materials
development}\label{tier-2-course-materials-development}

In Tier 2, an individual or a team of participants spent a semester to
develop Bayesian teaching materials supported with program organizers'
mentoring.

Tier 2 projects focused on what participants see as the greatest gap in
their current curriculum, taking into account the feasibility and the
space for innovation in their departments and programs. A project could
be designing a Bayesian course from scratch while another project could
be focusing on creating a Bayesian module with a particular scientific
application in a STEM field. A successful Tier 2 project could be
lecture notes, online interactive teaching tools, student-led learning
activities, among other things. Each Tier 2 project team shared the
teaching materials they developed, along with an overview video, at the
conclusion of the Tier 2.

All three authors served as a mentor in Tier 2. Each Tier 2 project team
had monthly check-in meetings with their mentor and all mentors held
monthly office hours throughout the semester. Every Tier 2 participant
received a stipend for their efforts.

All Tier 1 participants were eligible for Tier 2. An individual or a
team submitted an application, describing the teaching and learning
materials they would like to develop. The program organizers reviewed
the applications based on their feasibility, scope, and potential
impact. In both program cohorts, we aimed at selecting 10 participants
each year, and a total of 17 participants were selected for Tier 2.

\subsection{Tier 3: Dissemination}\label{tier-3-dissemination}

Tier 3 provided funding for selected Tier 2 projects to present their
products at conferences and/or through journal publications.

Tier 3 application was on a first-come-first-serve basis. Applicants
described the dissemination venue with their dissemination content and
plan and all three program organizers evaluated and made a joint
decision. Tier 3 award included a maximum dollar amount of support for
participants' dissemination effort. Across the two program cohorts, 6
participants have been supported through Tier 3 so far.

Next in Section~\ref{sec-outcomes}, we describe some demographic and
institution details of our recruited participants, and the impact they
created in their efforts in all three tiers.

\section{Program Outcomes}\label{sec-outcomes}

Over two program cohorts, we had a total of 35 participants attending
Tier 1. Of these, 17 participants continued onto Tier 2, and 6 continued
onto Tier 3. Participants were teaching at varying types of
institutions, including community colleges (\(n=\) 4), four-year
colleges (\(n=\) 9) and universities (\(n=\) 27). Particularly, five
participants were teaching at Hispanic Serving Institutions (HSIs) and
one was teaching at an Historically Black College or University (HBCU).

Table~\ref{tbl-disciplines} shows disciplines of departments of
participants from all cohorts and tiers. Evidently, our program
recruitment reached a wide range of STEM disciplines, a success of our
dedicated recruitment efforts. During Tier 1 bootcamp activities, such a
wide range of disciplines and backgrounds fostered many meaningful
exchanges in Bayesian applications in specific scientific settings, the
challenges faced by participants, and their innovative pedagogy
approaches.

\begin{ThreePartTable}
\begin{TableNotes}
\item \textit{Note: } 
\item Due to double departmental affiliations, the total number of participants listed in this table exceeds the total number of participants of the program.
\end{TableNotes}

\begin{longtable}[t]{lr}

\caption{\label{tbl-disciplines}Summary of scientific disciplines of
departments of program participants}

\tabularnewline

\\
\toprule
Discipline & Number of Participants\\
\midrule
Mathematics & 10\\
Statistics & 5\\
Biological sciences & 4\\
Computer science & 3\\
Anthropology & 2\\
\addlinespace
Business & 2\\
Engineering & 2\\
Physics & 2\\
Academic literacy and linguistics & 1\\
Economics & 1\\
\addlinespace
Education & 1\\
Integrated science and technology & 1\\
Law & 1\\
Political science & 1\\
Psychology & 1\\
\addlinespace
Sociology & 1\\
\bottomrule
\insertTableNotes

\end{longtable}

\end{ThreePartTable}

After the bootcamp in Tier 1, participants have modified their curricula
in different ways, regardless of Tier 2. Some have created modules to
implement Bayesian content by modifying their existing courses, some
have created stand-alone Bayesian courses, while others have created new
courses with a balance of frequentist and Bayesian approaches. Overall,
26 courses had directly been impacted by the program. Of these 7 were
newly created courses. One Tier 2 team even had their newly developed
Bayesian activity at one Accepted Student Days in Spring 2025 where
prospective students interested in Biology and Environmental Science sat
in on a mock class meeting.

Last but not least, we believe that program participants have taken
leadership in their disciplines to further the program impact by
disseminating Bayesian education content to their scientific
communities. Participants had delivered 9 presentations at state,
regional, national, and international conferences. A few more upcoming
presentations are scheduled. There are also some manuscripts in
preparation, while a few others submitted or under revision.

\section{Discussion}\label{sec-discussion}

As shown in Section~\ref{sec-outcomes}, our program has created learning
opportunities for undergraduate students in various STEM disciplines. In
addition, our participants have created impact in their own disciplines
through dissemination efforts. In this section, as we reflect on our
program as a TTT case study, we provide recommendations for
statisticians who might be interested leading TTT programs for STEM
educators. We note that while we believe that our program has served our
specific program aims well for the most part, there are different types
of programs that statisticians might lead. We end this section by
sharing some additional points that readers may consider for their own
TTT programs.

\subsection{Recommendations}\label{recommendations}

\textbf{Recommendation 1: Identify needs}

Developing a long-term TTT program requires a lot of financial and time
investment from program organizers, participants, and funding agencies.
A natural first step to take before such a commitment is to identify the
needs of target audience regardless of whether the TTT is at the local,
state, regional, national, or international level. To identify needs,
one can conduct interviews \citep[e.g.,][]{lee2005developing} and
surveys \citep[e.g.,][]{bennison2010learning}.

The approach we took was an extensive curricular analysis of Bayesian
courses at the national level \citep{DogucuHu2022TAS}. Through this
analysis, we saw that Bayesian courses are offered for undergraduate
majors other than statistics, mathematics, and data science, including
but not limited to biological sciences, economics, and public policy.
Considering the relevance of Bayesian methods across almost all science
disciplines, we tailored our TTT program to target all STEM instructors
teaching statistics. Our analysis also indicated that many existing
Bayesian courses have a number of prerequisites. Knowing that many
non-statistics, non-mathematics STEM majors do not take as many
mathematics or statistics courses, we tailored our TTT content to
include Bayesian topics with less prior statistics exposure, further
empowering our TTT participants to incorporate Bayesian methods in their
curricula in various forms and at differing levels.

In short, preliminary studies, interviews, and surveys to identify needs
can help TTT program organizers decide on program content and targeted
program audience.

\textbf{Recommendation 2: Build a community}

Community building is crucial for any TTT program so that it continues
to grow and flourish instead of being a one-time event. The idea of
connectedness reduces isolation and is a vital part of continuous
innovation \citep{eib2006faculty}. We were mindful about the fact that
our target audience was often the only faculty member or one of the few
focused on statistics methods in their departments and most likely the
only ones with exposure to Bayesian methods. Some of our participants
also worked at small colleges. Overall, the idea of planting a seed for
feeling connected to bigger scientific community and our Bayesian
education community is essential in our program. In addition to building
a sense of belonging, having a sense of community has helped
participants as well as program organizers learn from one another beyond
the program content such as teaching tips and tricks and tools.

Community building in a TTT can be achieved in many ways. We list a few
approaches that we have found effective in our program. As outlined in
Section~\ref{sec-overview}, our program built in social activities in
Tier 1 as part of our community building efforts. Online communication
platforms, such as Slack, were an effective tool to connect participants
with each other and with program organizers. We also fostered between
cohort interactions, such as inviting Tier 2 participants from a
previous cohort to join virtually during Tier 1 of the current cohort,
where successful Tier 2 projects were showcased. Many of our Tier 3
dissemination efforts were natural professional collaborations, such as
a program-themed organized session at a statistics conference and
manuscripts co-authored by our participants and their program organizer
mentors.

\textbf{Recommendation 3: Balance content knowledge and pedagogical
knowledge}

To be a successful statistics instructor, one's understanding of
statistical concepts is without a doubt vital, and knowing how to teach
statistics is just as important. The former is often referred to as
content knowledge and the latter as pedagogical knowledge. Needless to
say, knowing statistics and teaching statistics are not the same skill.
Both content knowledge and pedagogical knowledge contribute to teaching
performance \citep{shulman1987knowledge}. Thus, it is important for TTT
programs to deliver both content knowledge and pedagogical knowledge if
the target audience is educators.

In our program, we tried to deliver content knowledge by utilizing
existing textbooks and selecting topics deemed appropriate in the
literature \citep{DogucuHu2022TAS}. We delivered pedagogical knowledge
in multiple ways, including sharing examples from manuscripts focusing
on pedagogy, having participants envision a classroom activity, and
holding discussion on good and potentially bad pedagogical choices and
decisions every day during bootcamp in Tier 1. The three-tier program
design allowed such pedagogical knowledge to get distilled as
participants engage in Tier 2 when they developed their own course
materials. Many brought their own best and favorite pedagogical
practices into creating effective ways for their material delivery and
student learning in a classroom setting. Once in Tier 3, our
participants further disseminated the content knowledge and pedagogical
knowledge they have learned and practiced to a wider STEM audience.

\textbf{Recommendation 4: Get participants in charge of the change}

As mentioned in Section~\ref{sec-intro}, any TTT aims at creating change
through training the trainer. In our case, the change was to create more
exposure to Bayesian methods, especially at the undergraduate level.
Instructors play an important role in creating change in educational
contexts, especially in the curriculum. Even though all of our program
organizers have led, taught, managed, and mentored participants, in the
end the real change agents were the participants. As outlined in
Section~\ref{sec-outcomes}, the change our participants have created and
the impact brought by the change have taken place in their classrooms,
departments, institutions, and disciplines.

One effective strategy we had was to give participants full autonomy in
the type of change they want to create and how they want to create it.
We acknowledged and embraced the necessity for this level of
independence through understanding faculty as adult learners
\citep{eddy2019faculty}. For instance, despite having taught Bayesian
courses and content ourselves for many years, we as the program
organizers did not dictate a curriculum of our own for them to teach to
their students, nor specific pedagogy approaches they should take. We
instead supported the participants in whatever they envisioned for their
own students and other students in their disciplines, in terms of both
content and pedagogy. At the end of the day, it was the participants
themselves interacting with their students in the class and delivering
content that was designed and tailored for the specific STEM
disciplines. By supporting participants as the real change agents, we
fostered an active community that created tailored content with
innovative pedagogy approaches, exchanged ideas and best practices, and
disseminated useful content to the wider STEM education communities.

\textbf{Recommendation 5: Be ready to accommodate differences in
knowledge levels}

Anyone who teaches statistics or data science in the classroom knows
that each student comes with a unique background in statistics,
mathematics, and programming, among other things. Anyone who teaches a
TTT knows that this is true for students' trainers as well. Varying
backgrounds and knowledge levels are a real challenge for TTT programs
like ours because participants come from different scientific
disciplines, teach at different types of institutions (e.g., community
college vs.~four-year colleges), and teach different levels of courses
(e.g., introductory vs.~advanced). The challenge was to get participants
to a similar level of the planned content with limited time during the
bootcamp. Specifically in our case, it was to get every participant to
at least a basic level of understanding and teaching Bayesian methods.

We tried to overcome this challenge by providing asynchronous online
materials for on boarding prior to the bootcamp. In our case, it turned
out that this type of pre-bootcamp training helped us especially with
technological on boarding. All bootcamp materials utilized R
\citep{r2013r}, specifically the tidyverse packages
\citep{wickham2019welcome}. The materials were developed with Quarto,
and participants were encouraged to interact with Quarto documents such
as filling out their answers, taking notes, etc. Moreover, all materials
were shared through GitHub so participants needed at least a basic
understanding of Git and GitHub. Bringing all participants to a certain
level in use of all these tools was only possible through the
pre-bootcamp asynchronous videos and synchronous office hours, and our
program experience proved it to be an effective strategy.

We also took feedback from participants seriously in a mindful way to
improve our on boarding practice. For example, responding to the need
for a review of introductory probability and statistics content from the
first cohort, we incorporated such resources as part of the on boarding
process for the second cohort. This demonstrated that pre-bootcamp
training can be effective with content on boarding too.

\textbf{Recommendation 6: Teach additional good statistical practices
along the way}

Even though every TTT program has a focus topic, and in our case the
topic was Bayesian methods, participants get much more than just content
in a single topic. Therefore, we believe that when statisticians design
and deliver a TTT, we should deliver a whole set of good practices that
is relevant to the TTT.

For instance, in the program we utilized R, Quarto, Git, and GitHub, a
set of tools often used for reproducibility in research
\citep{erickson2024reproducible} and in the classroom
\citep{dogucu2024reproducibility, beckman2021implementing}. For an
instructor who had never used these tools, learning to use this set
could be overwhelming. The learning curve can especially be steep for
instructors who were programming for the first time. Nevertheless, we
took the route to provide a gentle introduction with guidance to teach
these good statistical practices before, during, and after the bootcamp,
instead of compromising and removing them from the program. In fact, we
received positive feedback from many program participants on learning
and using these tools as an important part of the training.
Nevertheless, even though participants understood the importance of the
set of tools, they may not not be equally comfortable with the
experience. We therefore recommend TTT program organizers to carefully
design the level and the breadth of good statistical practices they
intend to cover for their TTT program.

\subsection{Other considerations}\label{other-considerations}

There are a few additional considerations we would like to discuss that
reflect our program experience and could be helpful for statisticians
when designing and delivering a TTT program.

To cast a wide net of STEM disciplines in our recruitment efforts, we
inevitably ended up with participants who are the only ones in their
discipline (see Table~\ref{tbl-disciplines} in
Section~\ref{sec-outcomes}). In different programs, it might be possible
to have more then one participants from a STEM field in a given year to
create a field-specific community. If that cannot be achieved, then it
would be at least desirable to connect participants from the same
discipline across different training cohorts. This touches on the choice
between discipline-specific training versus discipline-agnostic
training. How to strike a balance in a TTT program which targets a wide
range of disciplines with limited training spots can be a challenge and
an opportunity, something we are still learning, exploring, and
experimenting.

While we put in deliberate effort in creating a community, its strength,
not surprisingly, was hard to maintain in the long term. The sense of
community was the strongest during the bootcamp when everyone met and
interacted with everyone else in person. After the bootcamp, members of
the same project team were able to stay connected while the connection
across teams was waning down. We certainly want to further our effort in
sustaining our community building among participants. At the same time,
we also hope to empower our participants to create and foster their own
communities in their specific STEM disciplines.

Last but not least, dissemination takes time to come to fruition, so it
is important to plan accordingly both in terms of funds and time. If
supporting participants' dissemination efforts is part of a TTT program,
then consider allowing a longer time frame for participants to find
venues to present their work. Needless to say, change takes time as
well, and under many constraints: participants' time and the room for
innovation in participants' departments and programs, to name a few. In
our TTT program, while some participants were able to introduce new
course materials as they were being developed as soon as the first
semester after the bootcamp, others are still waiting to be slated for a
specific course from their department's staffing plan to try out their
newly developed course materials. We look forward to hearing more
exciting updates from our program participants in the next few years. We
believe in patience when it comes to seeing real change and would
suggest the same for statisticians interested in designing and
delivering long-term TTT programs.

  \bibliography{bibliography.bib}

\end{document}